\documentclass[12pt]{article}

\advance\hoffset by -1.1truecm
\advance\voffset by -1.0truecm

\def\be{\begin{equation}}
\def\ee{\end{equation}}
\def\ba{\begin{eqnarray}}
\def\ea{\end{eqnarray}}
\newcommand{\Z}{\:\mbox{\sf Z} \hspace{-0.82em} \mbox{\sf Z}\,}

\newcommand{\R}{\mbox{I \hspace{-0.82em} R}}
\newcommand{\x}{{\bf x}}
\newcommand{\p}{{\bf p}}
\newcommand{\sn}{\smallskip\newline}
\newcommand{\mn}{\medskip\newline}
\newcommand{\bn}{\bigskip\newline}

\begin{document}
\title{Unsharp Degrees of Freedom and the \\ 
Generating of Symmetries}
\author{Achim Kempf\\
Institute for Fundamental Theory, Department of Physics\\
University of Florida, Gainesville, FL 32611, USA\\
{\small Email:  kempf@phys.ufl.edu}} 

\date{}

\maketitle

\vskip-7.5truecm

\hskip11.7truecm
{\tt UFIFT-HEP-99-08} 

\hskip11.7truecm
{\tt hep-th/9907160}
\vskip7.1truecm

\begin{abstract}
In quantum theory, real degrees of 
freedom are usually described by
operators which are self-adjoint. There are, 
however, exceptions to the rule.
This is because, in infinite dimensional Hilbert spaces, an 
operator is not necessarily self-adjoint even
if its expectation values are real.
Instead, the operator may
be merely symmetric. Such operators are not diagonalizable
- and as a consequence they 
describe real degrees of freedom which  
display a form of  ``unsharpness" or ``fuzzyness".
For example, there are indications that 
this type of operators could arise with the
description of space-time at the string
or at the Planck scale, where 
some form of unsharpness or fuzzyness 
has long been conjectured.

A priori, however, a potential problem 
with merely symmetric operators is
the fact that,  unlike self-adjoint operators, they  
do not generate unitaries - at least not straightforwardly.
Here, we show for a large class of these operators that they do
generate unitaries in a well defined way, and that these operators 
even generate the entire unitary group 
of the Hilbert space. This shows that 
merely symmetric operators, 
in addition to describing unsharp physical entities,
may indeed also play a  r{\^o}le in the generation of symmetries,
e.g. within a fundamental theory of quantum gravity.
\end{abstract}
\newpage

\tableofcontents
\newpage
\section{Introduction}
%%%%%%%%%%%%%%%%%%%%%%%%%%%%%%%%%%%%%%%%%%%%%%%%%
As a rule, 
real entities, or ``real degrees of freedom",
 are described in quantum theory 
through
operators which are self-adjoint.  There are, 
however, exceptions to the rule.
A common feature of these exceptional 
degrees of freedom is that they
display a form of ``unsharpness" or ``fuzzyness".
\mn
To see this, let us first recall that if an 
observable is described, as usual,  through a self-adjoint
operator, then the observable is 
absolutely sharp in the sense that,
in principle, it can be measured to arbitrarily 
fine resolution. This is because
 every self-adjoint operator $Q$ possesses 
 a spectral resolution, or
``eigenbasis". 
Indeed, if the system is in a 
$Q$-eigenstate, say $\vert q_n\rangle$, then the $Q$-uncertainty
\be
\Delta Q ({\vert\psi\rangle})~ :=~  \langle \psi\vert 
(Q-\langle \psi\vert Q\vert \psi\rangle)^2\vert \psi 
\rangle^{1/2}
\ee
(for $\vert \psi\rangle$ normalized) 
vanishes:
 \be
\Delta Q({\vert q_n\rangle})  =  \langle q_n\vert (Q-\langle
 q_n\vert Q\vert q_n\rangle)^2\vert q_n
 \rangle^{1/2}   =  0 
\ee
Also, 
eigenvectors to different eigenvalues are orthogonal.
Thus, if the system is localized with respect to 
the observable $Q$
around some value $q$ then the probability 
for finding it localized around
any other value $q^\prime$
 vanishes. Of course, if an ``eigenvalue"
  $q$ is in the continuous 
spectrum then the ``eigenstate" $\vert q\rangle$ 
is nonnormalizable and must be approximated by
a sequence of normalizable states. But also
 in this case,
the uncertainty $\Delta Q$ can be made arbitrarily
 small, and two eigenstates to different
points in the spectrum are
orthogonal, then with respect to the continuum 
normalization.
\sn
Thus, as is well-known, any self-adjoint 
observable $Q$ is ``sharp" in the sense that
its states of maximal $Q$-localization - the 
eigenstates $\{\vert q\rangle\}$ -
are orthogonal if localized around different 
values of $q$, 
and in the sense that each maximal localization 
state has vanishing uncertainty $\Delta Q$.
\bn
The reason why unsharp real degrees of freedom
 can occur is that, 
a priori, a real entity in a quantized theory
 may correspond to 
any operator whose expectation values are
 real - and, crucially, operators whose
expectation values are real need not be self-adjoint:
\sn
We recall that an operator whose expectation
 values are real is
a symmetric operator. In
finite dimensional Hilbert spaces, 
symmetric operators are 
self-adjoint and self-adjoint operators
 are symmetric. 
But in infinite dimensional Hilbert spaces,
 symmetric
operators need not be self-adjoint. This also 
means that 
they need not possess a spectral resolution
and diagonalization. As a consequence,
an operator $Q$ whose expectation values are real
 may also describe a
degree of freedom which 
is fuzzy or unsharp, in the sense that the minimum
 value for $\Delta Q$ may be
larger than zero, and$/$or those vectors which 
realize the minimum value for $\Delta Q$ may not be
orthogonal.
\sn
A simple example  in nonrelativistic 
quantum mechanics is
the momentum operator $\p$ of the particle
 in a box. For simplicity, let us consider
the case in one dimension, where the particle
 is confined, e.g.,
into the interval $[-L,L]$:
\sn
We recall that all
physical wave functions $\psi(x)$ vanish at the
 boundary of the box and
that the momentum operator $\p$ acts on
physical wave functions 
as the derivative operator
 $\p =-i\hbar\partial_x$. Indeed,  
$\p$ is an example of a merely symmetric operator: 
Clearly, all expectation values of $\p$ are real, and 
therefore $\p$ is symmetric.
On the other hand, $\p$ is not 
self-adjoint. This is because
plane waves do not vanish at the boundaries
and they are therefore not physical states.
This means that plane waves are not in 
the domain of $\p$, which implies that
$\p$ has no eigenbasis and no spectral resolution. 
\sn
As a consequence, $\p$ is indeed 
unsharp: From the uncertainty relation, since 
$\Delta x$ is surely smaller than $2L$,
we can expect that the minimum uncertainty 
in momentum is larger than zero. Indeed, the 
precise minimum value 
for $\Delta p$ is $\Delta p_{min}=\pi\hbar/2L$.
We will discuss this example in more detail below.
\mn
In general, for example
in any candidate theory for a fundamental 
theory of quantum gravity, 
it appears reasonable\footnote{
We will here not consider the alternative 
possibility of nonlinear operators.}   
to assume that entities which are described 
in the classical theory through real variables
are described in the quantized theory
through operators which are linear and whose 
expectation values are real.
Of course, we cannot assume that all those operators 
are observables in the usual quantum mechanical
 sense, nor that
they even act on the space of states. Instead, 
these operators may act 
on some Hilbert space of fields, or
branes (as we will briefly discuss below), or 
indeed on any abstract Hilbert space. 
\sn
On this level of generality we can only say that 
while some real degrees of freedom
may be described as self-adjoint operators, others
may be described by merely symmetric
operators - and that correspondingly the real
 physical entities which
they correspond to are  ``sharp" or ``unsharp". 
Interestingly, however, there exists
only a limited number of types of  sharpness or
 unsharpness, or ``short-distance structures",
which can occur with operators whose expectation 
values are real, i.e. with symmetric operators. 
A classification has been outlined in \cite{ak-erice}:
\sn 
Since the class of symmetric operators includes 
the self-adjoint operators, two of the
possible short distance structures are lattices 
and continua, corresponding to the fact that
self-adjoint operators occur with discrete and 
continuous spectra. The other extreme 
are the purely unsharp cases. These are described 
by the class of
simple symmetric operators. Those are operators
 which are symmetric but not self-adjoint, 
not even on any invariant subspace.
There exist subclasses of these operators which 
describe different types
of unsharpness. In \cite{ak-erice}, they
 have been divided into two broad classes, 
 fuzzy-A and fuzzy-B. Technically,
the two classes correspond the two possibilities 
of the deficiency indices being
equal or unequal.
\sn
In the present paper, we are concerned with those 
operators which describe entities that are unsharp 
of type fuzzy-A. Mathematically, these are the simple
 symmetric operators with equal deficiency indices.
For example, the momentum operator of the particle 
in a box is of the type fuzzy-A.
\sn
There are indeed theoretical indications that  
short-distance structures 
of the type fuzzy-A 
occur with the description of space-time at the Planck scale: 
\sn
Various theoretical arguments have long indicated that 
space-time displays a fundamental 
``foaminess", see \cite{hawking}, or unsharpness
 at very small lengths.
In particular, several studies, see e.g. 
\cite{gross}-\cite{haro}, suggest that
the structure of space-time at the Planck scale, or the 
string scale, is characterized effectively by
 correction terms to the uncertainty relations 
and, in particular, by corrections of the type:
\be 
\Delta X \Delta P ~\ge ~\frac{\hbar}{2} 
(1 + k~ (\Delta P)^2 +...)
\label{stucr}
\ee
As is easily verified, for any $k>0$,
 Eq.\ref{stucr} implies the existence of a 
 finite lower bound for $\Delta X$, namely:
 \be
 \Delta X_{min} =
\hbar \sqrt{k}
\ee
Here, $k$ is assumed to be a small 
positive constant
which is  related to the Planck scale, 
or in string theory to the string
scale. We here only 
remark that recent studies (on large extra 
dimensions that are seen by
gravity only) suggest that the unification 
and/or the Planck scale may even be 
as low as the TeV scale, see e.g.\cite{dienes}.
\sn
A positive minimum uncertainty $\Delta X_{min}>0$ 
arising from uncertainty relations of the
type of Eq.\ref{stucr} can be introduced as 
an ultraviolet cutoff in quantum field 
theories \cite{ak-jmp-qft,ak-gm}.
It has also been shown that this type of cutoff may 
solve the transplanckian energy paradox 
of black hole radiation, see
\cite{broutetal}. For general reviews of 
quantum gravity - 
and string theory motivations
of Eq.\ref{stucr}, see e.g. 
\cite{garay,witten}. For a recent discussion of
the potential origins of Eq.\ref{stucr} 
see e.g. \cite{santiago}, and
for a path integral approach to 
modified uncertainty relations see \cite{mangano}.
\sn
Technically, it is clear that any operator
 $X$ which obeys an uncertainty relation
of the type of Eq.\ref{stucr} cannot possess
 eigenvectors,
since the uncertainty $\Delta X$ would vanish 
for eigenvectors.
Therefore, any such operator $X$ can only be symmetric
but not self-adjoint. More precisely, it must be of
the type fuzzy-A, as was first shown in \cite{ak-erice}. 
We remark that operator realizations
and the functional analysis of uncertainty 
relations of the type of Eq.\ref{stucr} were first
discussed in \cite{ak-ucr}.
\bn
On the other hand, if we are to study those
cases in which a real degree of freedom 
is represented not
by a self-adjoint but instead by a merely 
symmetric operator, then
we must also address the fact that
 self-adjoint operators 
often play two r{\^o}les, namely both 
as real degrees of freedom \it and \rm
also as generators of symmetries.
Therefore,  the question arises, whether, 
or how, merely symmetric operators could
also be involved in the generating of symmetries.
Indeed, it is known that there is an important
 difference in this respect between self-adjoint and
merely symmetric operators. Namely,
merely symmetric operators, 
unlike self-adjoint operators, do not  generate unitaries,
at least not directly.
\mn
In the present paper, we therefore 
consider fuzzy-A type operators
with respect to the generation of 
unitary transformations - and 
we will find that these operators possess 
a remarkable property:
\mn
By definition, the fuzzy-A type operators 
are those operators $Q$ which on the physical
domain $D_Q=D_{phys}$ are simple symmetric 
with equal deficiency indices. 
For each such operator there exists a family of
operators $\{Q(\alpha)\}$ which 
coincide with $Q$ on the physical
 domain $D_Q$ and which 
are self-adjoint.
The $Q(\alpha)$ therefore generate unitaries
 in the usual way. 
We claim that these $Q(\alpha)$ - we 
recall that they
all coincide with $Q$ on the physical domain - 
generate, together,  \it all \rm unitary
 operators in the Hilbert space.
This shows that, in this way, 
operators of the type fuzzy-A can  indeed
relate to all aspects of symmetries in the 
Hilbert space on which they act.
\sn
We will also find that this result supports a 
 conjecture made in \cite{ak-erice,ak-euro}.
The conjecture proposes a mechanism by which 
those small wavelengths which are being cut off 
in the case of a fuzzy-A short-distance structure 
effectively turn
into internal degrees of freedom with an isospinor 
structure on which unitary groups act.

%%%%%%%%%%%%%%%%%%%%%%%%%%%%%%%%%%%%%%%%%%%%
\section{Examples of  Unsharp Degrees of Freedom}
Before we discuss the theorem, let us introduce
 concrete examples of simple symmetric operators
to which the theorem will apply.
\subsection{The momentum of the 
particle in a box}%%%%%%%%%%%%%%%%%%%%%%%
We have already mentioned the example of the 
 momentum operator $\p =-i\partial_x$ of 
 the particle in a 
box  (from now on we set $\hbar =1$). Since
 we will later use this example also
to illustrate the new theorem on generating 
symmetries,
let us discuss this case in more detail:
\sn
Assume the box to be the  one-dimensional 
interval $[-L,L]$. 
Due to the confining  box potential, all
physical wave functions $\psi(x)\in 
D_{phys}\subset H=L^2(-L,L)$ vanish at the boundary:
\be
\psi(-L)=0=\psi(L)
\label{bc1}
\ee
The expectation values of $\p$ are real:
\be
\langle \psi\vert \p \vert \psi\rangle \in \R,
 ~~~~~\mbox{for all ~~} \vert \psi\rangle \in D_{phys}
\ee
Thus, $\p$ is a symmetric operator. On the other hand, 
since no plane wave obeys the boundary condition,
 Eq.\ref{bc1},
$\p$ does not possess (normalizable nor
nonnormalizable) eigenvectors. Thus $\p$ is not 
self-adjoint, instead $\p$ is simple symmetric.
\sn
Even though there are no plane waves among the
 physical states,
plane waves can of course be
approximated by sequences of physical states which
 are 
approximately plane waves within most of the 
interval $[-L,L]$, but which also quickly
decay to zero towards the boundaries, such as to
 always obey the boundary condition, Eq.\ref{bc1}.
\sn
One may therefore be tempted to assume that $\p$ 
is still ``approximately" self-adjoint and should
therefore describe a sharp entity. This is, 
however, not the case: Indeed, as we already mentioned, 
$\p$ is unsharp in the sense that for all physical 
states $\vert \psi\rangle \in D_{phys}$ the
momentum uncertainty is bounded from below by a
 fixed finite amount:
\be
\Delta p(\psi) ~\ge ~\Delta p_{min} = \frac{\pi}{2L} 
~~~~~\mbox{for all normalized}
 ~~\vert\psi\rangle~ \in~ D_{phys}  
\ee
Intuitively, the reason is that the larger the 
part of the interval on which a physical 
wave function approximates a plane wave, the 
steeper it must decay to zero towards the
boundaries. The steep decay necessarily yields
 a significant contribution to the action
of the derivative operator $\p$. This is 
connected to the fact 
that $\p$ is a noncontinuous
operator. We remark that only noncontinuous i.e. only 
unbounded operators can be simple symmetric and 
display this unsharpness. For an explanation 
of the unsharpness phenomena
in these terms, see \cite{ak-euro,ak-bialo}.
\sn
Here, let us explicitly calculate 
the physical states with the lowest momentum uncertainty.
To this end, we solve the variational problem of minimizing 
\be
(\Delta p)^2 = \langle \psi\vert \p^2\vert\psi\rangle
 -\langle\psi\vert \p\vert\psi\rangle^2
\ee
by minimizing $\langle\psi\vert\p^2\vert\psi\rangle$ 
under the constraints 
$\langle\psi\vert\p\vert\psi\rangle =\rho$ and 
$\langle\psi\vert\psi\rangle=1$, and the boundary
condition Eq.\ref{bc1}.
\sn
Introducing Lagrange multipliers $k_1,k_2$, the 
functional to be minimized is:
\be
S=\int_{-L}^L dx~\left\{-(\partial_x\psi^*)(\partial_x\psi) + 
k_1 (\psi^*\psi - c_1) + k_2 (-i
 \psi^*\partial_x\psi-c_2)\right\},
\ee
yielding the Euler-Lagrange equation:
\be
\partial_x^2 \psi + k_1 \psi -i\partial_x\psi ~=~0
\ee
For each choice of momentum expectation 
value $\langle \p\rangle =\rho$, there is  
(up to a phase) one normalized and the boundary
 condition obeying solution:
\be
\psi_\rho(x)~=~ L^{-1/2}~\cos\left(\frac{\pi x}{2
 L}\right)~e^{i \rho x}
\label{tenn}
\ee
These are the physical wave functions which minimize 
$\Delta p$. We see that the $\psi_\rho(x)$
are essentially plane waves, apart from the 
modulus, which
approaches zero at the boundaries, as it must,
 being a physical state.
It is clear that the modulus of the wave 
functions $\psi_\rho(x)$ goes to zero with just 
the optimal steepness to minimize the momentum 
uncertainty $\Delta p$.
\sn
The minimum value for the uncertainty in the momentum 
of a particle in the box
is now readily calculated from the 
solutions $\psi_\rho$, as:
\be
\Delta p_{min} ~=~ \sqrt{\langle \psi_\rho\vert 
\p^2\vert \psi_\rho\rangle - \rho^2}
~=~ \frac{\pi}{2 L}
\ee
In this case here, the minimum uncertainty 
$\Delta p_{min}$ does not depend 
on the expectation value $\rho$. Note that
 for generic simple symmetric operators with
equal deficiency indices the
minimum standard deviation can depend on the 
expectation value:
\be
\Delta Q_{min} = \Delta Q_{min}(\langle Q\rangle).
\ee
It is standard procedure to verify that the deficiency 
indices of $\p$ are indeed equal, 
namely $(1,1)$. Thus, the
short-distance structure (of momentum space) is of
 the type fuzzy-A in the
terminology of \cite{ak-erice}.

\subsection{An ``unsharp" position
 operator}%%%%%%%%%%%%%%%%%%%%%%%%%%%%
Let us now illustrate the same phenomena
 with the example of a simple symmetric  
operator which is given explicitly in terms
of an infinite dimensional matrix.
\sn
Consider the operator $Q$ which is defined 
as the matrix
\be
Q = 
\left(
\matrix{
0 & a_1 &  0 & 0 &  \ldots \cr
                           a_1 & 0 & a_2 & 0 & 
                            \ldots \cr
                           0 & a_2 & 0 & a_3 & 
                            \ldots \cr
                           0 & 0 & a_3 & 0 & 
                            \ldots \cr
                           \vdots & \vdots &
                            \vdots &\vdots & \ddots 
} \right)
\ee
where we define the matrix elements $a_n$ through 
\be
a_n~:= ~\sqrt{1+s+s^2+...~s^{n-1}}
\ee
with $s$ being a constant, obeying $s\ge 1$. 
Of course, one may use the partial geometric series:
\be
1+s+s^2+...~s^{n-1}~=~(s^n-1)/(s-1).
\ee
We define the domain  $D_Q$ of $Q$ to consist 
of all column vectors which possess an 
arbitrary but finite number of nonzero entries. 
The domain $D_Q$ is dense in the Hilbert
 space $H=l^2$ of all 
of square summable vectors, i.e. $\overline{D_Q}=H$.\sn
Clearly, 
\be Q_{ij} = Q_{ji}^{*}. \ee
Thus, on its domain, $D_Q$, the 
expectation values of  $Q$ are real, i.e.
$Q$ is a symmetric operator.
\mn
Let us consider first the special case $s=1$. 
\sn
In the case $s=1$, the matrix elements 
reduce to $a_n=\sqrt{n}$. We recognize that $Q$ is then 
the ordinary essentially self-adjoint 
quantum mechanical position operator, in its 
Fock space representation.
Its spectrum is the real line and there exist 
sequences of vectors in its domain $D_Q$
such that $\Delta Q$ becomes arbitrarily small,
 i.e. $Q$ is a sharp observable.
\mn
The situation is qualitatively different for  $s>1$.
\sn
It has been shown in \cite{ak-ucr} that if $s>1$,
 then for all vectors in $D_Q$ the uncertainty
$\Delta Q$ is finitely bounded from below, 
by:
 \be \Delta Q_{min}= \sqrt{1-s^{-1}} \ee
This means that
for all normalized $\vert \phi \rangle \in D_Q$, i.e. 
for all normalized vectors with an arbitrary but
finite number of nonzero entries, the
 uncertainty in $Q$ obeys:
\be 
\Delta Q({\vert \phi \rangle}) ~=~ \langle \phi \vert (Q-\langle 
\phi \vert Q \vert \phi\rangle)^2\vert
 \phi \rangle^{1/2} ~\ge~ \Delta Q_{min}.
\ee
The fact that $\Delta Q_{min} = \sqrt{1-s^{-1}}$
 is larger than zero implies that   
there are no (normalizable nor nonnormalizable) 
eigenvectors of $Q$, i.e. that
 $Q$ is not self-adjoint. $Q$ has been shown to 
 be simple symmetric of type fuzzy-A,
 see \cite{ak-erice,ak-ucr}.
\sn
In fact, in this example, the minimum standard 
deviation is a nontrivial function of
the expectation value of $Q$. For simplicity, we 
have only given  the absolute minimum
$\Delta Q_{min}$. For the precise form of 
$\Delta Q_{min}(\langle Q\rangle)$
and its derivation, see \cite{ak-ucr}.
\mn
Finally, we remark that no finite dimensional
 truncation of the matrix  $Q$ could 
possess a nonzero minimum uncertainty 
$\Delta Q_{min}>0$. This is because
the notions of symmetry and  self-adjointness
 only differ
on infinite dimensional Hilbert spaces:
Every finite dimensional symmetric matrix is also
self-adjoint and therefore possesses 
eigenvectors $\vert q\rangle$
for which, of course, $\Delta Q(\vert q\rangle) =0$.   
\subsection{Unsharpness from noncommutativity}%*********
The statement that self-adjoint
 operators can \it always \rm 
be resolved to arbitrary precision is 
compatible with the Heisenberg uncertainty principle.
Assume, for example, that $S$ and $T$ are 
two self-adjoint observables which do
not commute.
\sn
Then, there  holds the uncertainty relation:
\be
\Delta S~\Delta T~ \ge ~ 
\frac{1}{2}~\vert\langle[S,T]\rangle\vert
\ee
This implies of course that 
if $\Delta T$ is smaller than some value, 
say $\Delta T < t_0$ and if, say, 
 $[S,T]=i 1$ then the RHS is nonvanishing, yielding 
$\Delta S \ge 1/2t_0$. Thus, in this case, $\Delta S$
cannot be made arbitrarily small but possesses instead
a finite lower bound $\Delta S_{min}=1/2t_0$.
\sn
This is not a contradiction to the statement 
that self-adjoint operators can 
always be diagonalized, because
to require $\Delta T \le t_0$ is to restrict
 the Hilbert space to only those
states for which $\Delta T \le t_0$ holds.
On this restricted domain, 
the operator $S$ is not self-adjoint, instead 
it is simple symmetric.
\sn
In general, noncommutativity of symmetric operators
 in any physical theory  
induces an interplay between the domains of those 
operators, which in turn
 affects  whether or not they are self-adjoint or 
 merely symmetric. Indeed, 
even more generally, not only kinematical but also 
dynamical operator equations, i.e.
not only commutation relations but also operator
 equations of motion can affect the 
domains of operators, and can therefore affect 
whether or not 
these operators are symmetric or self-adjoint. 
\sn
Thus, while it is a well-known and 
much-discussed phenomenon 
that the sharpness or unsharpness of 
real entities in quantum theory 
can depend on the \it kinematics \rm - through
uncertainty relations - it appears that 
there is a priori no reason 
to exclude the possibility that 
the sharpness of real entities can also 
change \it dynamically, \rm 
for example in a fundamental theory of quantum gravity.

%%%%%%%%%%%%%%%%%%%%%%%%%%%%%%%%%%%%%%%%%%%%%%%%%%%%%
\section{Unsharp Degrees of Freedom and 
the Generating of Unitaries}
As is well-known,
self-adjoint operators often act not only as real
degrees of freedom, but simultaneously also 
as generators of symmetries.
Merely symmetric operators, on the other hand, 
do not directly generate unitary operators.
\sn
This appears to indicate that while symmetric
 operators possess 
the interesting property of being able to 
describe unsharp 
real degrees of freedom, 
they should not be able to play a r{\^o}le in
 the generation of symmetries.
\sn
Here, we will therefore address the problem of
 the generation of unitary operators
for the class of simple symmetric operators
 which describe fuzzy-A type short-distance 
 structures, i.e. which have
 equal deficiency indices. This class includes  
our examples above, and it includes, in particular, all 
operators $X$ with a finite lower bound 
$\Delta X_{min}>0$. 
\sn
We will prove the following: 
\sn
For each simple symmetric operator $X$ with 
equal deficiency indices, acting
on a physical domain $D_{phys}$ which is dense 
in a Hilbert space $H$,
there exists a family of self-adjoint operators
 $X(\alpha)$ which coincide with $X$ on the
physical domain. We claim that these
 operators $X(\alpha)$,
together, generate
the full unitary group of the Hilbert space.
This result shows that, in this way,
the operators of this class can relate to all aspects
of symmetry in the Hilbert space on which they act.  
\sn
The precise formulation of the general 
theorem and its proof are given in 
Sec.\ref{s-five}. 
Before, however, we will give a detailed illustration
of the theorem in concrete examples.

%%%%%%%%%%%%%%%%%%%%%%%%%%%%%%%%%%%%%%%%%%%%
\subsection{The theorem in concrete examples}
\label{s-four}
In order to demonstrate the mechanism by 
which simple symmetric operators are able to 
generate all unitaries of the Hilbert space, 
let us consider a concrete example  
in ordinary nonrelativistic quantum mechanics 
in one dimension.
\sn
In this case, for the particle on the real line, 
the operators $\x$ and $\p$ are
self-adjoint and can be exponentiated to yield
 unitaries:
\mn
We may represent the operators $\x$ and $\p$, 
irreducibly, as the self-adjoint
multiplication and 
differentiation operators $\x.\psi({{x}}) = 
{{x}}\psi({{x}})$ and $\p.\psi({{x}})
= -i\partial_{{x}}\psi({{x}})$ acting 
on a dense domain in the Hilbert space $H$ of 
square integrable
wave functions $\psi({{x}})$ over the real line.
\sn
As is well-known,
$\x$ and $\p$, \it together, \rm  generate \it all
 \rm unitary operators $U$ on the
Hilbert space $H$, via the Weyl formula
\be
U = \int\int \frac{ds dt}{2\pi \hbar} ~u(s,t)~ 
\exp[i (s\x +t\p)/\hbar]  \label{weyl}
\ee
where the $u(s,t)$ are suitable complex-valued 
functions. In fact, all bounded operators
$B \in B(H)$ can be generated in this way.
\mn
On the other hand, we can also represent $\x$ 
and $\p$ reducibly, for example, as
the self-adjoint multiplication and 
differentiation operators 
\be
\x.\psi_i({{x}}) ~=~ {{x}}\psi_i({{x}}) 
~~~\mbox{and}~~~ \p.\psi_i({{x}})
= -i\partial_{{x}}\psi_i({{x}})
\ee
acting on a Hilbert space of wave functions 
$\psi_i({{x}})$ on the real line which possess
an additional ``isospinor" index, running $i=1,...,n$. 
\sn
The  scalar product of wave functions then 
contains an iso-sum:
\be
\langle \psi \vert \phi\rangle = 
\sum_{i=1}^n\int_{-\infty}^{\infty}
 d{{x}} ~\psi_i^*({{x}})\phi_i({{x}})
\ee
Clearly, $\x$ and $\p$ are acting diagonally 
in the isospinor space. Therefore,
$\x$ and $\p$ do not generate the $U(n)$ of
 the isorotations.
Thus, in this case,
the Weyl formula, Eq.\ref{weyl}, does not 
yield all bounded
operators nor does it yield only all 
unitaries on the Hilbert space. 
Only if we supplemented the operators 
$\x$ and $\p$  by 
additional hermitean $n\times n$ matrices, 
$T_i$, could we generate
$U(n)$ on the isospinor space
and therefore all of $B(H)$. 
\mn
Let us now consider again the case where the
particle is confined to the interval $[-L,L]$. 
As we saw above,   
the momentum operator $\p=-i\partial_{{x}}$
is then no longer self-adjoint and it is instead simple 
symmetric of type fuzzy-A. Therefore, $\p$
then matches  the conditions of our proposition.
\mn
Namely, our proposition is that for any simple
 symmetric operator $\p$ of type fuzzy-A, 
e.g. the momentum of the particle in a box,
there exists a one-parameter family of 
self-adjoint operators $\p(\alpha), 
~~(0\le \alpha<2\pi)$,~ such that:
\sn
\begin{itemize}
\item each $\p(\alpha)$ coincides with $\p$ on the
 physical domain, i.e. $$\p(\alpha)\vert \psi 
\rangle = \p \vert \psi \rangle~~~~~\mbox{
 for all}~~~ \vert \psi\rangle \in D_{phys}$$ 
\item the $\p(\alpha)$, together, (weakly) 
generate the algebra $B(H)$ of bounded operators
on the Hilbert space, which includes of course 
the full unitary group on $H$. 
\end{itemize}
$ $
\sn
Indeed, we claim that, unlike in the Weyl formula,
 the operator $\x$ is now no longer needed
to generate $B(H)$, because
the operators $\p(\alpha)$ alone
already generate $B(H)$ (even though each $p(\alpha)$
coincides with $\p$ on the dense physical domain $D_{\p}$).
\sn
Even further, we can consider the case 
where the wave functions of the particle
in the box carry some
isospinor index. Then, $\p$ is again simple
 symmetric of type fuzzy-A and 
our theorem  applies. We claim that there 
then exists a multi-parameter 
set of self-adjoint operators 
 $\p(u)$, which again all coincide with $\p$ on
physical states, and which generate all of $B(H)$!
This means that there is no need to introduce 
isospin rotation generators $T_j$ by hand,
since the $\p(u)$ are able to generate all: 
translations, phase rotations and isorotations.
\subsection{Generating $B(H)$ in the scalar 
case}%****************************************
To see this, we consider first the case without 
an isospinor index. 
\sn
As discussed above, the 
momentum operator $\p$, acting as $\p.\psi({{x}})
 = -i \partial_{{x}}\psi({{x}})$
on the physical wave functions $\psi\in D_{phys}$ 
over the interval is a simple symmetric operator.
We recall  that although all physical wave functions 
vanish at the boundary, 
they are a dense set in the Hilbert space of square 
integrables $\overline{D_{phys}}=H=L^2(-L,L)$.
\mn
Let us now construct a family of operators $\p(\alpha)$ which 
coincide with $\p$ on the physical domain $D_{phys}$, 
but whose domain is larger and
who are self-adjoint on this larger domain.
\sn
To this end, we define the operators 
$\p(\alpha)$  by extending
the domain $D_{phys}$ such as to include
 wave functions  which 
are periodic up to a phase
\be \label{bc} 
\psi(-L) = e^{i \alpha} \psi(L) \label{boundc},
\ee
where $\alpha$ is some arbitrary but fixed real number.
To be precise, the domain of the self-adjoint
 extension $\p(\alpha)$ is therefore
\be
D_{\p(\alpha)} ~:=~ D_{phys} ~\cup~ 
\left\{\psi({{x}})\in D_{\p^*}
 ~\vert~\psi(-L)=e^{i\alpha}\psi(L) \right\},
\ee
where $D_{\p^*}$ is the domain of
 the adjoint operator $\p^*$.
\sn
Note that $e^{i\alpha}$ must be a fixed phase in order 
to ensure that the boundary terms cancel 
in the partial integrations which are
needed to show that 
$\langle \psi_1 \vert (\p(\alpha)\vert \psi_2\rangle) = 
(\langle \psi_1 \vert \p(\alpha)) \vert \psi_2\rangle$.  
\sn
Indeed, for 
each fixed choice of a
 phase $e^{i\alpha}$ there exist 
eigenvectors of $\p(\alpha)$, i.e.
plane waves, $\psi_n^{(\alpha)}(x)$,
which obey the corresponding boundary condition:
\be
\psi_n^{(\alpha)}(x)~ =~ e^{i\omega_n x} 
~~~~~~~\mbox{where}~~\omega_n=
\frac{2\pi n-\alpha}{2L}, ~~~n\in \Z
\ee
As is straightforward to check, 
the $\psi_n^{(\alpha)}$ form an orthonormal 
eigenbasis of $\p(\alpha)$, and
each $\p(\alpha)$ is self-adjoint.
\mn
Let us now consider the implications for the 
generating of unitaries:
\mn
If the wave functions were not restricted to 
the interval, $\p$ would be self-adjoint and
$\p$ could be exponentiated 
to obtain a unitary operator, say \be U(a)~:= 
~\exp(i a \p),\ee for some $a\ge 0$. The
action of this unitary is to translate 
wave functions  by the amount $a$ to the right: 
\be U(a).\psi({{x}}) ~=~e^{a\partial_{{x}}}
~\psi({{x}})
~=~
 \psi({{x}} +a).
\label{dfg}
\ee 
In the case where the particle is confined
 to the box, however, 
i.e. where the
Hilbert space only consists of wave functions 
on the interval, the operator 
$\p$ is not self-adjoint and 
 cannot be exponentiated: The formal expression
 $U(a) =  \exp(i a \p)$ is now not a unitary 
 transformation, because it would translate 
beyond the interval boundaries, which is not 
defined in the Hilbert space.
\sn
Nevertheless, for the particle in a box, 
there exists, as we saw, a whole family of 
self-adjoint extensions $\p(\alpha)$
of $\p$. Since each  $\p(\alpha)$ is self-adjoint, 
each can be exponentiated and the resulting
operator 
\be
U_\alpha(a) := \exp(i a \p(\alpha)) 
\ee
is unitary. The action of $U_\alpha(a)$ on 
wave functions is again to translate
wave functions to the right (for $a>0$),
 as in Eq.\ref{dfg}. Now, however, due to the 
 boundary condition,
Eq.\ref{bc}, the part of the wave function which 
would be translated beyond the right interval
boundary reappears into the interval from the 
left, with the same modulus, but phase shifted by
the phase $e^{i\alpha}$.   
\sn
Thus, the unitary $U_\alpha(a)$  translates the 
wave functions by the
amount $a$ and phase shifts the
wave functions by $e^{i\alpha}$ when translating 
them beyond a 
boundary and into the interval again from the 
opposite boundary. 
\sn
Let us consider the composition of such unitaries. 
Crucially, the product 
\be
U_{\alpha^\prime}(-a)U_{\alpha}(a)
\ee
 is a unitary operator which does
not translate wave functions. This is  because
the first factor translates by $a$ and the second
 factor translates back by the same
 amount. Nevertheless, 
since the two factors translate with different 
phase shifts,
the product
is not the identity operator. Namely,
 $U_{\alpha^\prime}(-a)U_{\alpha}(a)$ is the 
 unitary operator whose action is to 
leave the modulus of wave functions 
unchanged, but to
phase shift the wave functions
on a part of the interval. 
\sn
E.g., choosing some $a\in[0,2L]$, the action is
\sn
\be
U_{\alpha^\prime}(-a)~U_{\alpha}(a). \psi({{x}}) ~=~
 \left\{\matrix{~~\psi({{x}}),  & 
 ~~~\mbox{for}~~& x\in[-L,L-a] \cr \cr
e^{i(\alpha-\alpha^\prime)} \psi({{x}}), 
 & ~~~\mbox{for}~~& x\in [L-a,L]}\right. 
\ee
$$ $$
By suitable composition of operators $U_\alpha(a)$ 
for various $a$ and $\alpha$ 
it is therefore  possible to 
generate unitaries which yield \it arbitrary \rm 
local phase rotations of wave functions.
\sn
For example, choosing some $a,b$ obeying  
$0< b<a<2L$, we form the operator:
$$ $$
\be 
U_{\alpha}(-(a-b))~U_{0}(-b)~
U_{\alpha}(a). \psi({{x}}) ~~~~~~~~~~~~~~~~~~
~~~~~~~~~~~~~~~~~~~~~~~~~~~
~~~~~~~~~~~~~~~\mbox{$ $ }
\ee 
$$ $$
$$ \mbox{ }~ ~~~~~~~~~~~~~~~~=~
\left\{\matrix{\psi({{x}}), & ~~\mbox{for}~& 
x\in [-L,L-a] \cup 
[L-a+b,L] \cr \cr
e^{i(\alpha)} \psi({{x}}), & ~~\mbox{for}~ & 
x\in [L-a,L-a+b]}\right. ~~~~~~~
$$
$$ $$
The action of this operator
 is to phase rotate wave functions by $e^{i\alpha}$
in the interval $[L-a,L-a+b]$  and to leave 
the wave functions
invariant outside that interval.
\sn
Thus, remarkably, the set of self-adjoints 
which coincide with $\p$
on the physical domain is able to generate all 
translations \it and \rm also all local phase rotations,
while we recall that in the case where $\p$ is
 self-adjoint, 
the operator $\x$ is needed order to generate 
phase rotations, namely
through $e^{i f(\x)}\psi(x)=e^{if({{x}})}\psi(x)$.
\subsection{The case with isospin}%************
We consider again
a particle constrained to the interval $[-L,L]$. 
The particle's wave function $\psi_i({{x}})$
 shall now carry an
isospinor index $i=1,...,n$. 
The scalar product in the Hilbert space of square 
integrables on the interval then includes an iso-sum:
\be
\langle \psi \vert \phi\rangle = 
\sum_{i=1}^n\int_{-L}^Ld{{x}} ~\psi_i^*({{x}})\phi_i({{x}})
\ee
Due to the box potential, the physical wave functions, 
$\vert \psi \rangle \in D_{phys}$, again 
obey the boundary condition \be \psi_i(-L) 
=0=\psi_i(L),~ ~~~~(i=1,...n)\ee
The action of $\p$ is diagonal in iso-space: 
$\p.\psi_i({{x}}) = -i\partial_{{x}}\psi_i({{x}})$. 
Again, there are no plane waves in the
physical domain and therefore the momentum 
operator on the physical domain 
 is not self-adjoint. Instead, $\p$ is simple 
 symmetric (with deficiency indices $(n,n)$).
Self-adjoint extensions $\p(u)$ are now
obtained by enlarging the domain of 
$\p$ to include wave functions
which obey the boundary condition
\be
\psi_i(-L) ~=~ \sum_{i=1}^n~u_{ij}~ \psi_j(L)
\label{bcis}
\ee
where $u_{ij}$ is any unitary 
$n \times n$ matrix, generalizing the 
 phase $e^{i\alpha}$ of the scalar case above.
 As is readily checked, the proof of self-adjointness
of the $\p(u)$ requires again the cancellation of 
the boundary terms which arise through 
the partial integrations needed to show that 
$(\langle \psi\vert \p(u))\vert \phi\rangle = 
\langle \psi\vert( \p(u)\vert \phi\rangle)$,
and this cancellation is achieved exactly
by the boundary conditions of the form of Eq.\ref{bcis}.
\sn
As in the scalar case, 
while $\p$ does not directly yield
 unitaries, each of the self-adjoint 
$\p(u)$ which reduce to $\p$ on the 
physical
domain does generate unitaries, e.g.  
by exponentiation ($a$ real) :
\be
U_u(a):= e^{i a p(u)}
\ee
The unitaries $U_u(a)$ again act on wave functions 
by translating them by the amount $a$, and,
due to the self-adjoint extensions' boundary conditions, 
any part of 
the wave function which hits a boundary 
reappears from the other side into the interval, 
now iso-rotated
by the matrix $u$ (or by $u^{-1}$ if $a$ is negative). 
\sn
It is possible to proceed as in the scalar
 case, composing such unitaries
to translate the wave functions 
back and forth, using different self-adjoint
 extensions.
It is clear that in this way arbitrary local 
isorotations can be generated. 
\sn
Thus, the set of self-adjoint operators which 
reduce to $\p$ on the
physical domain indeed generates not only 
translations, which they may be expected to,
 but also arbitrary
local phase rotations, and - if an isospinor 
index is present - then
they even generate all local iso-rotations.
\mn
We now proceed to the proof of the theorem 
for the general case. 
\section{Theorem}%*****************
\label{s-five}
\subsection{Definitions}%%%%%%%%%%%%%%%%%%%%%%  

Let us recall that a symmetric operator
 $X$ is called \it simple symmetric 
\rm if $X$ is not self-adjoint and if it possesses 
no invariant subspace such that the 
restriction of $X$ to this subspace 
yields a self-adjoint operator. Our
 examples above  are simple symmetric. 
\sn
Further, we recall that the  \it Cayley
 transformed \rm operator $S$ of 
a symmetric operator $X$, defined as
\be
S:= (X-i1)(X+i1)^{-1}
\ee
is isometric. An isometric operator is called
 \it simple isometric \rm 
if it cannot be reduced to an
invariant subspace such that the reduced
 operator is unitary.
It is known that a subspace 
reduces a symmetric operator $X$ if and only
 if it reduces its  Cayley transform, see,
for example, \cite{AG}. Note, however, that
 not every isometric operator is the Cayley
transform of a symmetric operator.
\subsection{Theorem} %%%%%%%%%%%%%%%%%%%%%%%
Let $X$ be a closed simple symmetric operator 
with equal deficiency indices, defined on a 
domain ${\bf D}_X$ which is 
dense in a complex Hilbert space ${\bf H}$. 
Then, the self-adjoint extensions 
$X(\alpha)$ of $X$
generate a $^*$-algebra ${{\cal A}}$ 
which is weakly dense in ${\cal{B}}({\bf H})$.
Thus, in particular, the self-adjoint 
extensions generate the full unitary group
${\cal{U}}({\bf H})$ of the Hilbert space. 
\subsection{Outline of the proof}%%%%%%%%%%%%%%
The first step will be to use 
the $X(\alpha)$ to generate 
a suitable set ${{\cal M}}$ of unitaries, 
which in turn generate an algebra ${\cal A}$.
The proof then consists 
in showing that the commutant ${{\cal A}}'$ 
of the algebra ${{\cal A}}$ is
${{\cal A}}' = \mbox{\bf{C}}1$. This implies 
that its double commutant is
${{\cal A}}'' = {\cal{B}}({\bf H})$. The 
proposition then follows since, with v. Neumann, 
the double commutant of any $^*$-algebra 
is its weak closure.
\subsection{Proof}%%%%%%%%%%%%%%%%%%%%%%%%%%%%
We begin by choosing a suitable set of unitaries which are
generated by the self-adjoint 
extensions $X(\alpha)$ of $X$.
To this end, consider 
the isometric Cayley transform $S$ 
of $X$ \be S := (X-i1)(X+i1)^{-1}\ee
with domain \be
{\bf D}_S = (X+i1).{\bf D}_X .\ee
We define the \it local group \rm ${\cal{T}}$ 
as the set of all unitaries which map the deficiency 
space ${\bf D}_S^\perp = ((X+i1).{\bf D}_X)^\perp$ 
onto itself and which act as the 
identity on ${\bf D}_S$, i.e.:
\be {{\cal T}} := \{T ~\vert~ T: {\bf D}_S 
\rightarrow {\bf D}_S, ~ T: {\bf D}_S^\perp \rightarrow  
{\bf D}_S^\perp, ~T_{\vert {\bf D}_S} =
 1, ~TT^\dagger = T^\dagger T = 1  \} . \ee 
It is clear that the local group, ${{\cal T}}$, 
is isomorphic to the
unitary group $U(n)$, where $n$ is the deficiency
 index $n := \mbox{dim}({\bf D}_S^\perp)$.
\mn
Since, by assumption, both deficiency 
indices are equal, i.e. both 
spaces \be L_\pm = ((X\pm i1).{\bf D}_X)^\perp\ee are
of equal dimension, there exist unitary extensions of $S$. 
\sn
Let $U$ be one of the unitary extensions of $S$:
\be 
U^\dagger U = U U^\dagger =1, ~~
U: L_+ \rightarrow L_-, ~~
U_{\vert {\bf D}_S} = S.\ee
We consider now
the coset \be 
{{\cal M}}:= \{M ~\vert ~~ M = U T,
 ~~T \in {{\cal T}}\}\ee
of  unitary extensions of $S$.
\sn   
Indeed, as is well known, each unitary 
extension
of the Cayley transform $S$ of a symmetric
 $X$, i.e. here 
each element of ${{\cal M}}$, is indeed 
generated, via the Cayley transform, by a
self-adjoint extension $X(\alpha)$ of $X$. 
\sn
We will now show that the $^*$-algebra 
${{\cal A}}$ generated by ${{\cal M}}$ 
is weakly
dense in ${\cal{B}}({\bf H})$. As mentioned, 
this follows from v. Neumann's double commutant
theorem if we can prove that only multiples 
of the identity operator
commute with ${\cal{M}}$, i.e. with $U$ and 
all elements 
of ${{\cal T}}$.
\mn
To this end, let us consider an operator 
$V$ which obeys: \be 
\vert\vert V \vert\vert < \infty ~~~ \mbox{and} ~~~
[V,U]=0=[V,T],~~ \forall~ T\in {{\cal T}} \label{two} \ee
We need to show that $V$ is a multiple 
of the identity operator.
\mn
Since the closure of $X$ implies the closure
 of the deficiency space ${\bf D}_S^\perp$ 
 and of ${\bf D}_S$,
we can use  ${\bf H}= {\bf D}_S \oplus 
{\bf D}_S^\perp$ to write $V$ and the elements $T \in
{{\cal T}}$ in block form:
\be
T = \left( \matrix{1 & 0 \cr 0 & t} \right), 
 ~~~~~~~~~
V = \left( \matrix{V_{{{\bf D}_S}{\bf D}_S} & 
V_{{{\bf D}_S}{\bf D}_S^\perp} \cr
V_{{{\bf D}_S^\perp}{\bf D}_S} & V_{{{\bf 
D}_S^\perp}{\bf D}_S^\perp}} \right)
\ee
Here, $t= T_{\vert {\bf D}_S^\perp}$, i.e. $t:~
 {\bf D}_S^\perp ~\rightarrow~ 
 {\bf D}_S^\perp$ and, 
e.g.,  $V_{{{\bf D}_S}{\bf D}_S^\perp}: 
~{{\bf D}_S^\perp}~ \rightarrow ~{\bf D}_S$. 
In this notation,
$[T,V]=0$ reads:
\be
\left( \matrix{ 0 &  V_{{{\bf D}_S}{\bf
 D}_S^\perp} (1-t)\cr \cr
(t -1) V_{{{\bf D}_S^\perp}{\bf D}_S} & [t, 
V_{{{\bf D}_S^\perp}{\bf D}_S^\perp}]} \right)
~~=~~0 \label{bm}  
\ee  
\sn
Eq.\ref{bm} holds for all $T \in {{\cal T}}$, 
and
in particular it holds for unitaries $t: 
{\bf D}_S^\perp \rightarrow {\bf D}_S^\perp$ for which 
the value $1$ is a regular point, e.g. $t =-1$.
Thus, $V_{{{\bf D}_S}{\bf D}_S^\perp} = 0$ 
and $V_{{{\bf D}_S^\perp}{\bf D}_S} =0$.
\sn
Further, ${{\cal T}}$ is the full unitary group on
${\bf D}_S^\perp$. It is therefore 
irreducibly represented on ${\bf D}_S^\perp$. 
Thus, $[t, V_{{{\bf D}_S^\perp}{\bf
 D}_S^\perp}]=0, ~\forall t$
 implies with Schur that
$V$ acts on ${\bf D}_S^\perp$ as a 
multiple of the identity, i.e. 
$V_{{{\bf D}_S^\perp}{\bf D}_S^\perp} =
 {\lambda} 1$ where ${\lambda} \in \bf{C}$.
In block matrix form, $V$ therefore reads:
\be
V = \left( \matrix{V_{{{\bf 
D}_S}{\bf D}_S} & 0 \cr
0 & {\lambda} 1} \right)
\ee
Consider now the kernel \be
{\bf K}:= \mbox{ker}(V-{\lambda} 1).\ee 
By construction, ${\bf D}_S^\perp \subset {\bf K}$,
and ${\bf K}^\perp \subset {\bf D}_S$.
As the kernel of a closed operator, ${\bf K}$ is closed. 
We wish to show that in fact
 ${\bf K}={\bf H}$ and ${\bf K}^\perp =\emptyset$,
which is to say that $V= {\lambda} 1$. 
\sn
To this end, let us assume the opposite,
 namely that ${\bf K}^\perp \neq \emptyset$.
\sn 
We can then use ${\bf H}= {\bf K}^\perp 
\oplus {\bf K}$ to write
both $V$ and $U$ in a new block form:
\sn
\be
V = \left( \matrix{V_{{\bf K}^\perp 
{\bf K}^\perp} & 0 \cr
V_{{\bf K} {\bf K}^\perp} & {\lambda} 1} 
\right), ~~~~~~~~~~~
U = \left( \matrix{U_{{\bf K}^\perp 
{\bf K}^\perp} & U_{{\bf K}^\perp {\bf K}} \cr
U_{{\bf K} {\bf K}^\perp} & U_{{\bf K} {\bf K}}} \right)
\ee
$$ $$
The relation $[V,U]=0$ now reads:
\sn
\be
\left( \matrix{... ~~ , & (V_{{\bf K}^\perp 
{\bf K}^\perp} - {\lambda} 1) U_{{\bf
 K}^\perp {\bf K}} \cr \cr
... ~~,  & V_{{\bf K} {\bf K}^\perp} 
U_{{\bf K}^\perp {\bf K}}  } \right) ~~~= ~~~ 
\left( \matrix{0~~ , & 0~~ \cr \cr
0 ~~ ,  & 0 ~~  } \right) \label{ten}
\ee
$$ $$
On the other hand, $U_{{\bf K}^\perp 
{\bf K}}.{\bf K} \subset {\bf K}^\perp$, i.e. the 
range of $U_{{\bf K}^\perp {\bf K}}$ 
is not in the kernel of the operator $(V-{\lambda} 1)$:
\sn
\be
(V-{\lambda} 1)\vert w\rangle  = \left( 
\matrix{(V_{{\bf K}^\perp {\bf K}^\perp}-{\lambda}
 1)\vert w\rangle\cr \cr
V_{{\bf K} {\bf K}^\perp}\vert w\rangle}\right)  
       ~~         \neq ~~0 ,
~~~~\forall ~\vert\omega\rangle \neq 0, \vert 
\omega\rangle \in   U_{{\bf K}^\perp {\bf K}}.{\bf K}
\label{el}
\ee
\sn
Thus, the existence of any nonzero vector 
$\vert w\rangle \in {\bf K}^\perp$ in the range 
$U_{{\bf K}^\perp {\bf K}}.{\bf K}$ would contradict   
Eq.\ref{ten}. Consequently, 
the range of $U_{{\bf K}^\perp {\bf K}}.{\bf K}$
 is empty, i.e. $U_{{\bf K}^\perp {\bf K}}=0$.
\sn
Therefore,  ${\bf K}$ is an invariant
subspace for $U$. Since also $[U^{-1},V]=0$,
 it follows analogously 
that ${\bf K}$ is an invariant subspace for
$U^{-1}$. Thus, ${\bf K}$  and ${\bf K}^\perp$ 
both reduce $U$:
\sn
\be
U = \left( \matrix{U_{{\bf 
K}^\perp {\bf K}^\perp} & 0 \cr
0 & U_{{\bf K} {\bf K}}} \right)
\ee
$$ $$
Since  $U_{\vert {\bf D}_S} = S$ and ${\bf 
K}^\perp \subset {\bf D}_S$
we have $U_{{\bf K}^\perp {\bf K}^\perp} = 
S_{{\bf K}^\perp {\bf K}^\perp}$.
This implies that ${\bf K}^\perp$
 is an 
invariant subspace for $S$, on which $S$ is unitary. 
However, the simplicity of $X$ implies 
that also $S$ is simple, i.e. 
$S$ does not have any invariant subspace on which it
would be unitary.
\mn
Thus, in fact, ${\bf K}^\perp = \emptyset$ 
and ${\bf K}={\bf H}$. Consequently, 
$V={\lambda} 1$, which had to be shown.
\mn
With von Neumann this implies that
the weak closure of the $*$-algebra ${{\cal A}}$ 
generated by $1, U$ and the elements of
${{\cal T}}$ is the algebra ${\cal{B}}({\bf H})$ 
of all bounded operators on the Hilbert space, 
and ${\cal{B}}({\bf H})$ includes of course all unitaries.
We recall that this means that 
for each  bounded operator $B \in {\cal{B}}({\bf H})$
 there exist sequences of operators 
$B_n \in {{\cal A}}$ such that $$\lim_{n\rightarrow \infty}
 \langle \psi \vert B - B_n \vert \phi \rangle = 0
~~~~\forall~~ \vert \psi\rangle, ~
 \vert \phi \rangle \in H.$$
\sn
Thus, for any simple symmetric $X$ with 
equal deficiency indices the
set of self-adjoint operators which coincide 
with $X$ on its domain
generate indeed (e.g. via generating the coset ${\cal M}$)
the full unitary group of the Hilbert space.
\subsection{A corollary}
%**************************
As we mentioned before, in finite dimensional 
Hilbert spaces every symmetric operator,
i.e.  every operator
whose expectation values are real, i.e. every
 matrix obeying $X_{ij}=X_{ji}^*$, is 
also self-adjoint. Therefore, in finite 
dimensional Hilbert spaces, there are 
no simple symmetric operators, i.e. our 
theorem cannot be applied.
\sn
Let us add, however, that the above proof 
yields as a corollary that 
any simple isometric operator with equal 
deficiency indices has the property that
its unitary extensions, together, generate 
all unitaries and ${\cal B}(H)$. 
And indeed,  there exist simple isometric 
operators also in finite dimensional Hilbert spaces.
\sn
As an illustration, let us consider the 
simple case of the 
two dimensional Hilbert space spanned by 
normalized vectors $e_1,e_2$.
We define a linear operator, $S$, as the map 
which maps 
$S: e_1 \rightarrow e_2$. Clearly, $S$ is not 
unitary, because of its
limited domain $D_S := {\bf C}e_1$ and range 
${\bf C} e_2$. Also, $S$ does not have any invariant
proper subspace. $S$ is norm preseving where 
it is defined. Thus, $S$ is  
 a simple isometric
operator. The dimensions of its deficiency spaces,
 i.e. of the orthogonal complements of its
domain and range are both 1, i.e. they are equal. 
Thus, $S$ is an operator to which the corollary
of our theorem applies. The claim is that the
 unitary extensions of $S$ 
generate all $2 \times 2$ matrices,
including of course the unitaries.
\sn
To see this, we begin by choosing one unitary
 extension $U$ of $S$, e.g.:
\be
U~=~ \left(\matrix{0~~1\cr 1~~0}\right)
\ee
The elements $T(\alpha)$ of the 
local group ${\cal T}$ of all unitaries which 
act as the identity on $D_S$ and which act as
a unitary on $D_S^\perp$ are of the form
\be
T(\alpha) ~:=~ \left(\matrix{1 & 0\cr 0 & 
e^{i\alpha}}\right)
\ee
where $e^{i\alpha}$ is any arbitrary phase. 
Thus, each unitary extension of $S$ is of the 
form $U~T(\alpha)$ for some $\alpha$. Indeed, 
the algebra generated by $1,U$ and the
unitary extensions $T(\alpha)$ is all of
 $M_2({\bf C})$, as is clear because it contains
for example the Pauli matrices:
\be
\sigma_1~ =~ U,~~~~~~\sigma_2 ~= ~i ~U~ 
T(\pi),~~~~~~ \sigma_3 ~= ~T(\pi).
\ee 
On the other hand, we recall that simple 
symmetric operators only exist in infinite dimensional
Hilbert spaces. 
Indeed, in our 2-dimensional example here, 
the inverse Cayley transform 
$X$ of $S$ does exist, 
\be
X ~= ~ i~(S+1)(S-1)^{-1} ~=~ 
-i\left(\matrix{1~~0\cr 2 ~~1}\right)
\ee
but $X$ is clearly not symmetric.
In general, the inverse Cayley 
transform of a simple isometric 
operator is not necessarily a simple symmetric 
operator. In particular, it cannot be simple 
symmetric in
finite dimensional Hilbert spaces. On the other
 hand, the fact that, vice versa, the
Cayley transform of a simple symmetric operator 
is always a simple isometric operator is what 
we used in the theorem.
\section{Conclusions and Outlook}%**************
Our subject of investigation is any physical entity, 
within a quantized theory, which is real in the
 sense that it is described by an operator $Q$
whose expectation values are real.
\mn
Our first conclusion has been that any such real 
entity or ``real degree of freedom" 
can only be ``sharp" or ``unsharp" in a few 
well-defined ways. 
Namely, the physical entity is sharp if the 
operator $Q$ is self-adjoint.
In this case, its possible short-distance 
structures are lattices and continua.
On the other hand, the physical entity is 
unsharp if  the operator $Q$ is merely symmetric. 
Then, its possible short-distance structures are what 
we call fuzzy-A and fuzzy-B. All other 
possibilities are mixtures of these.
The sharpness or unsharpness of a real entity
 can depend on the
kinematics of the theory, e.g. through commutation 
and uncertainty relations. A priori,
the unsharpness of a real entity can also be a
 function of the dynamics, e.g. through 
 operator equations of motion.
Several properties of the fuzzy-A and fuzzy-B 
short-distance structures are
discussed in \cite{ak-erice}. A more detailed 
classification is in preparation.
\mn
Secondly, and this has been the main 
subject of the present work, we considered that
self-adjoint operators often not only 
represent real degrees of freedom
but that they can also act as generators of symmetries. 
We therefore investigated in which way also
operators that describe fuzzy 
degrees of freedom could generate symmetries. 
\mn
To this end, we focussed on the 
class of operators of the type fuzzy-A. We
 found that these possess a 
remarkable property:
If, on the physical domain, $D_{phys}$, an 
operator $Q$ is of the type fuzzy-A (i.e. 
simple symmetric with equal deficiency indices),
then there exists a set of self-adjoint 
operators $\{Q(\alpha)\}$ (the self-adjoint extensions)
which all agree with $Q$ on the physical 
domain. We showed that
the operators $Q(\alpha)$, together, 
generate all unitaries and all
bounded operators in the Hilbert space.  Thus, 
in this way, at least the fuzzy operators
of type fuzzy-A can indeed play a r{\^o}le in
 all aspects of
symmetries in the Hilbert space in which they act.
\mn
In our investigation of the properties of
physical entities which are described by operators
 whose expectation values are real,
we did not make any assumptions about the 
interpretation of these operators, nor about the
underlying physical theory. 
Therefore, our conclusions, 
firstly about the possible types of sharp- and 
unsharpness
and secondly about  these operator's ability to 
generate symmetries,
apply to all 
linear operators which describe real 
degrees of freedom - for example
in candidate theories
for a fundamental theory of quantum gravity.
\mn
Indeed, for example the matrix model for
 M-theory, see e.g.
\cite{polchinski}, does employ symmetric 
operators, $X_i$, to encode
space-time information. In this case, the 
matrix elements of the $X_i$ are interpreted
 in terms of coordinates of $D0$-branes. 
Initially, the $X_i$ are finite
 dimensional, say $N\times N$ matrices.
The quantization and the necessary limit 
$N\rightarrow \infty$ are highly 
nontrivial, but it is clear that
the resulting operators will still 
be at least symmetric.
The short-distance structure which they describe 
will therefore fall into the 
classification outlined in \cite{ak-erice}.
The $X_i$ are in general 
 noncommutative, which is of course 
a kinematical source for fuzzyness, but there may also 
exist dynamical causes for fuzzyness of the $X_i$.
It is clear that if, or when, these operators
are of type fuzzy-A, then our
present results show how they relate 
to the unitary group of
the Hilbert space on which they act.
\mn
Studies in the context of quantum groups,
 see e.g. \cite{majid-pl,majid},
and in the wider field of 
noncommutative geometry, have yielded
new approaches to building
models for space-time at the Planck scale,
see e.g. \cite{connes}-\cite{cc}. Some of 
this work has been shown to be related
to string theory, see 
e.g. \cite{nc}. As far as these models of space-time 
apply linear operators to describe real entities 
we are covering these operators.
It should be very
interesting to investigate the 
r{\^o}le of the present results in this context.
\mn
On the other hand, even on the level of generality
 on which we have been working here, 
more conclusions can likely be drawn:
Indeed, one of the examples which we gave for our theorem 
indicates a particular direction for further investigation:
\sn
We discussed the case of the simple symmetric
 differential operator $\p=-i\delta_{ij}\partial_x$ 
which acts on a domain of wave 
functions $\psi_i(x)$ with an isospinor index
$i=1,...,n$, defined over the interval $[-L,L]$.
There exists a whole $U(n)$- family of
 self-adjoint operators $\p(u)$ 
which coincide with $\p$ on its domain. 
We showed that, even though $\p$ itself acts
diagonally on the isospinor space, the 
operators $\p(u)$
are able to generate all unitaries in the 
Hilbert space - which includes, in particular,
also all isospinor rotations.
\sn
We can interpret this result as providing an
example for a  conjecture made in 
\cite{ak-erice,ak-euro}. The conjecture
is that simple symmetric operators with
 equal deficiency indices $(n,n)$
\it always  induce \rm 
isospinor structures of dimension equal 
to the deficiency index.\sn
The conjecture also yields an intuitive
 physical interpretation of the effect of an
ultraviolet cutoff of the type fuzzy-A: 
Namely, those short wavelengths which
are being cut off at short distances  are 
being turned
effectively into internal degrees of freedom
 associated with a unitary isospinor structure.
This would mean that local gauge group structures
 could have their
origin in what we here called the local unitary 
group on the deficiency spaces.
Work in this direction is in progress.
\bn
\bf  Acknowledgement: \rm 
The author is happy to thank John Klauder 
for useful criticisms.

\end{document}